\documentclass[a4paper,12pt]{article}
\usepackage{amssymb,amsmath,amsfonts}
\usepackage{fullpage}
\usepackage[latin1]{inputenc}
\usepackage{graphicx}

\title{The Weibull-Geometric Distribution}
\author{Wagner Barreto-Souza$^{\rm a}$, Alice Lemos de Morais$^{\rm a}$ and Gauss M. Cordeiro$^{\rm b}$
\\\\
$^{\rm a}$Departamento de Estat\' \i stica,\\ 
Universidade Federal de Pernambuco,\\
Cidade Universitária, 50740-540 -- Recife, PE, Brazil\\
(e-mail: wagnerbs85@hotmail.com, alice.lm@hotmail.com)
\\\\
$^{\rm b}$Departamento de Estat\'{\i}stica e Inform\'atica,\\
Universidade Federal Rural de Pernambuco,\\
Rua Dom Manoel de Medeiros s/n, 50171-900 -- Recife, PE, Brazil\\
(e-mail: gausscordeiro@uol.com.br)
\\\\
}
\date{}

\begin{document}
\maketitle
\begin{abstract}
In this paper we introduce, for the first time, the
Weibull-Geometric distribution which ge\-ne\-ralizes the
exponential-geometric distribution proposed by Adamidis and Loukas
(1998). The hazard function of the last distribution is monotone
decreasing but the hazard function of the new distribution can take
more ge\-ne\-ral forms. Unlike the Weibull distribution, the
proposed distribution is useful for modeling unimodal failure rates.
We derive the cumulative distribution and hazard functions, the
density of the order statistics and calculate expressions for its
moments and for the moments of the order statistics. We give
expressions for the Rényi and Shannon entropies. The maximum
likelihood estimation procedure is discussed and an algorithm EM
(Dempster et al., 1977; McLachlan and Krishnan, 1997) is provided
for estimating the parameters. We obtain the information matrix and
discuss inference. Applications to real data sets are given to show
the flexibility and potentiality of the proposed
distribution.\\

{\bf keywords}: EM algorithm; Exponential distribution; Geometric
distribution; Hazard function; Information matrix; Maximum
likelihood estimation; Weibull distribution
\end{abstract}

\section{Introduction}

Several distributions have been proposed in the literature to model
lifetime data. Adamidis and Loukas (1998) introduced the
two-parameter exponential-geometric (EG) distribution with
decreasing failure rate.  Kus (2007) introduced the
exponential-Poisson distribution (following the same idea of the EG
distribution) with decreasing failure rate and discussed various of
its properties. Marshall and Olkin (1997) presented a method for
adding a parameter to a family of distributions with application to
the exponential and Weibull families. Adamidis et al. (2005)
proposed the extended exponential-geometric (EEG) distribution which
generalizes the EG distribution and discussed various of its
statistical properties along with its reliability features. The
hazard function of the EEG
distribution can be monotone decreasing, increasing or constant.\\

The Weibull distribution is one of the most commonly used lifetime
distribution in modeling lifetime data. In practice, it has been
shown to be very flexible in modeling various types of lifetime
distributions with monotone failure rates but it is not useful for
modeling the bathtub shaped and the unimodal failure rates which are
common in reliability and biological studies. In this paper we
introduce a Weibull-geometric (WG) distribution which generalizes
the EG and Weibull distributions and study some of its properties.
The paper is organized as follows. In Section 2, we define the WG
distribution and plot its probability density function (pdf). In
Section 3, we give some properties of the new distribution. We
obtain the cumulative distribution function (cdf), survivor and
hazard functions and the pdf of the order statistics. We also give
expressions for its moments and for the moments of the order
statistics. The estimation by maximum likelihood using the algorithm
EM is studied in Section 4 and inference is discussed in Section 5.
Illustrative examples based on real data are given in Section 6.
Finally, Section 7 concludes the paper.

\section{The WG distribution}

The EG distribution (Adamidis and Loukas, 1998) can be obtained by
compounding an exponential with a geometric distribution. In fact,
if $X$ follows an exponential distribution with parameter $\beta Z$,
where $Z$ is a geometric variable with parameter $p$, then $X$ has
the EG distribution with parameters $(\beta,p)$. Since the Weibull
distribution generalizes the exponential distribution, it is natural
to extend the EG distribution by replacing in the above compounding
mechanism the exponential by the Weibull distribution.

\begin{figure}[h!]
\centering
\includegraphics[width=0.50\textwidth]{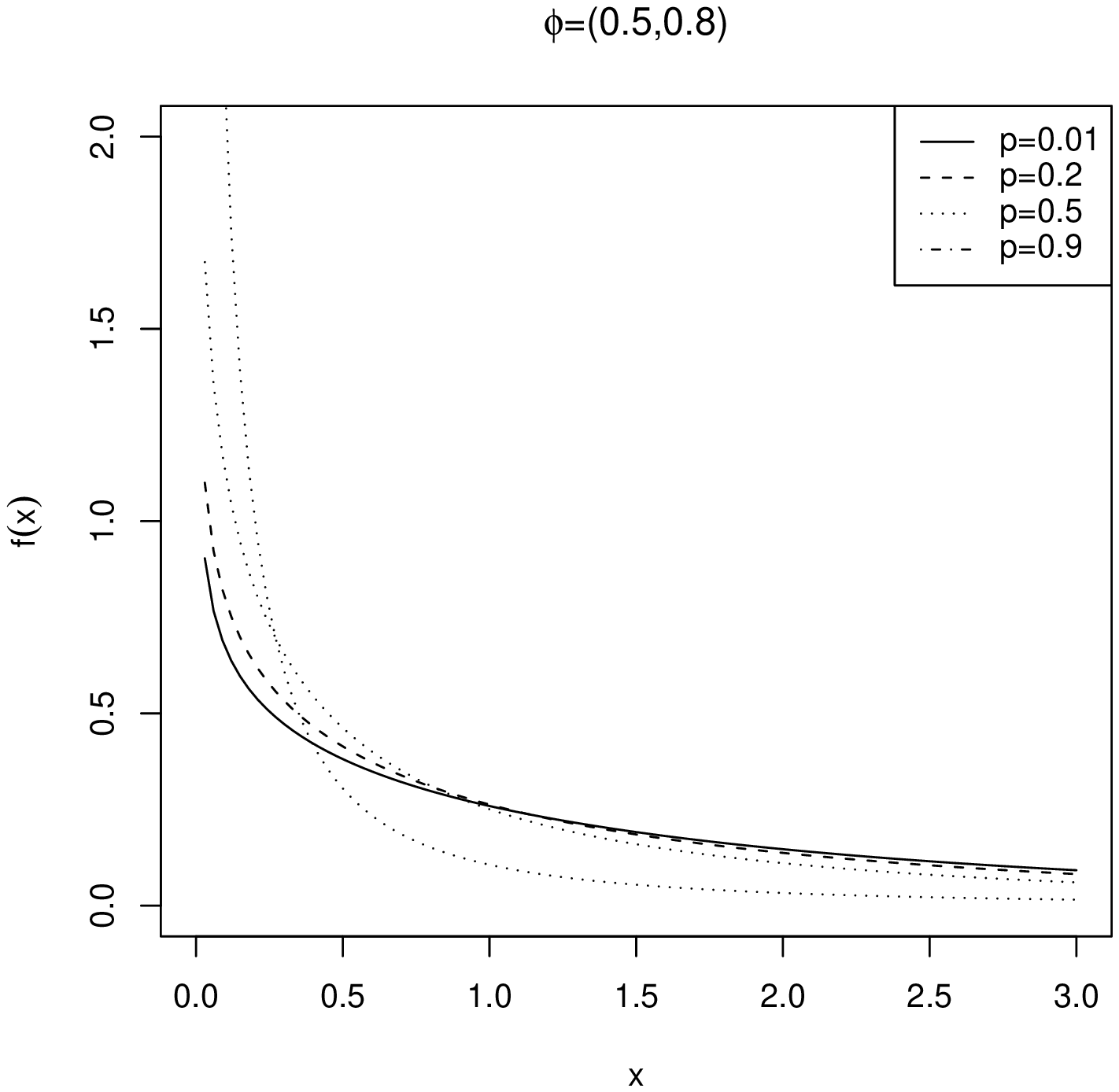}\includegraphics[width=0.50\textwidth]{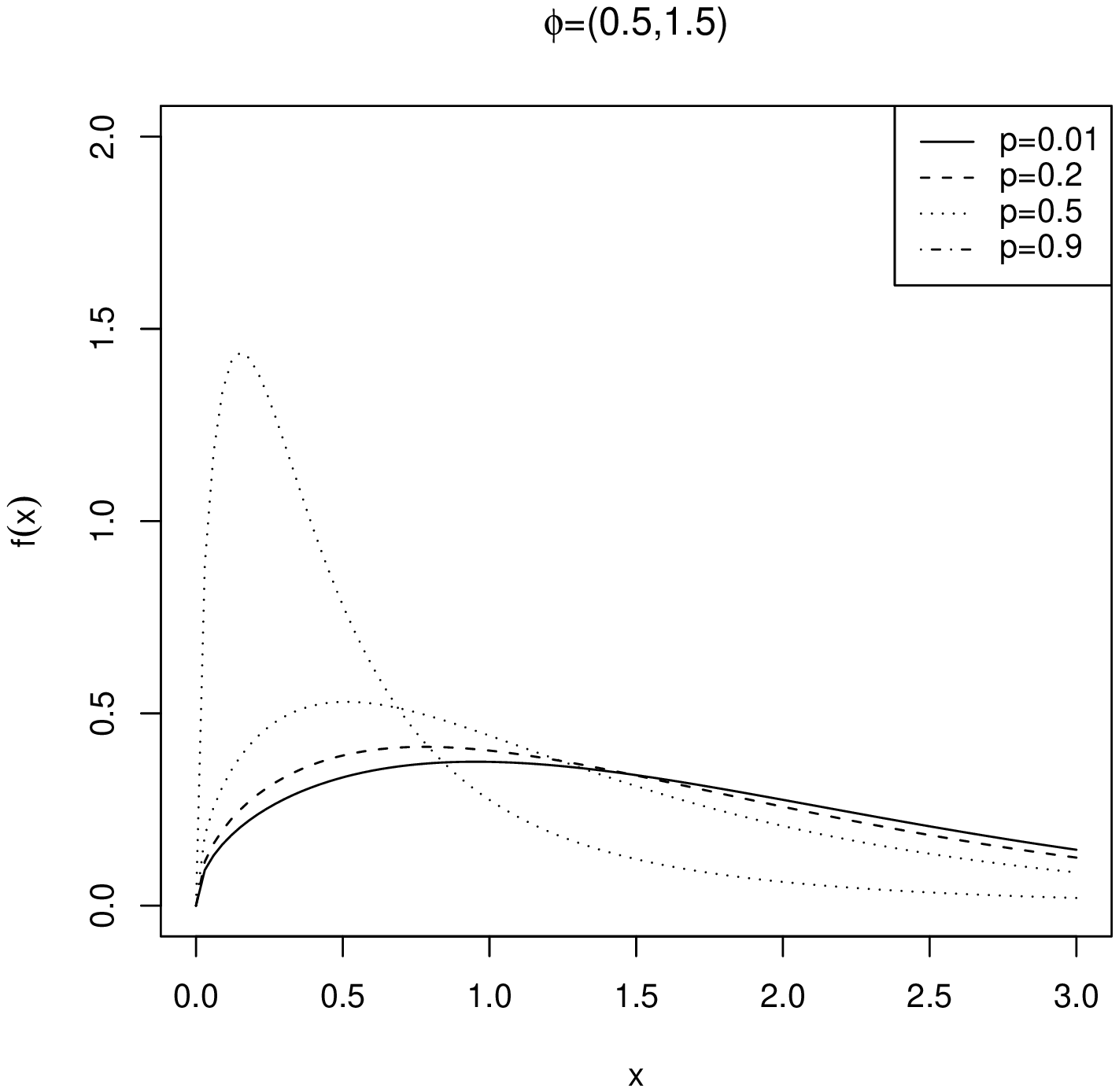}
\includegraphics[width=0.5\textwidth]{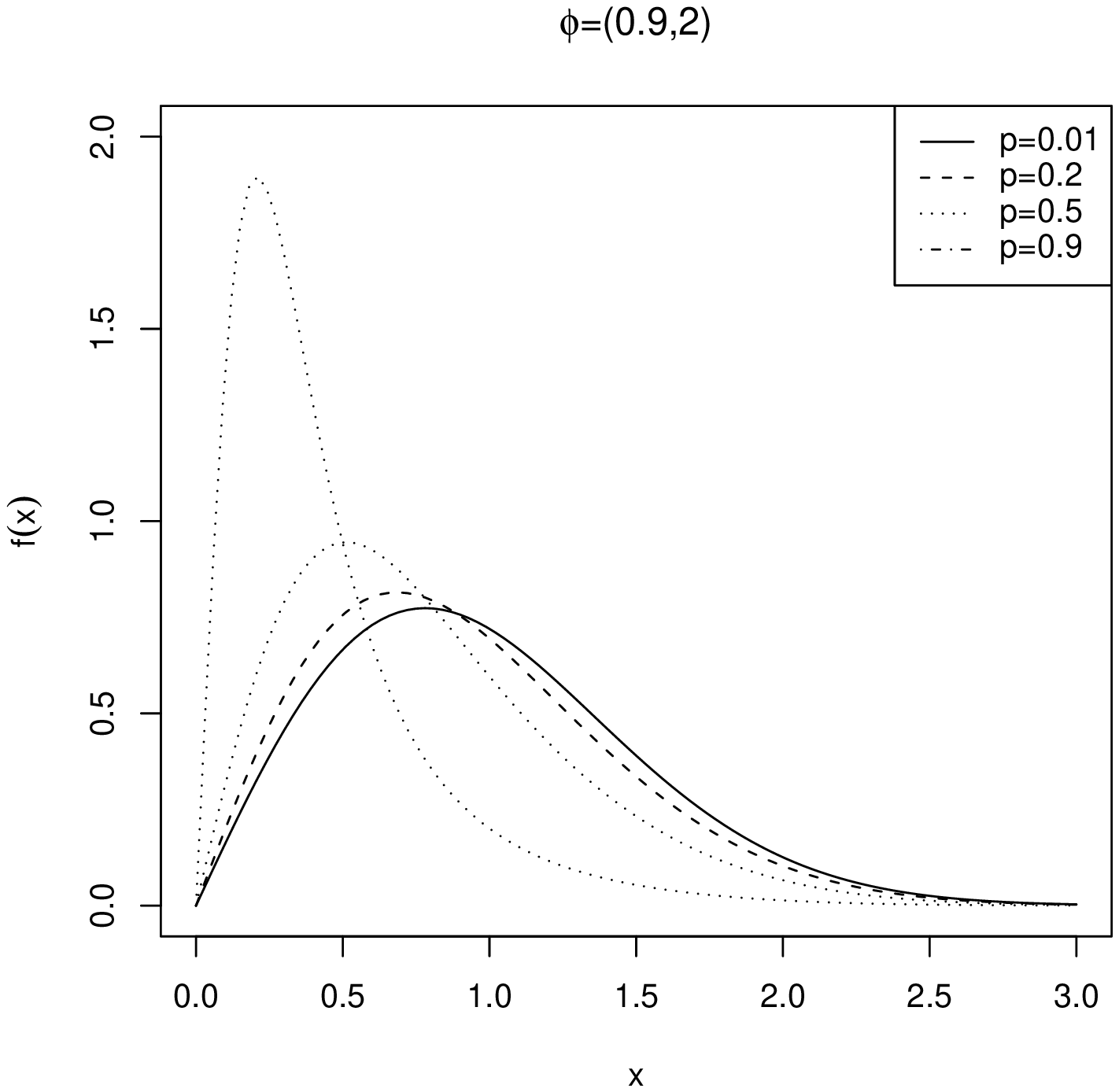}\includegraphics[width=0.50\textwidth]{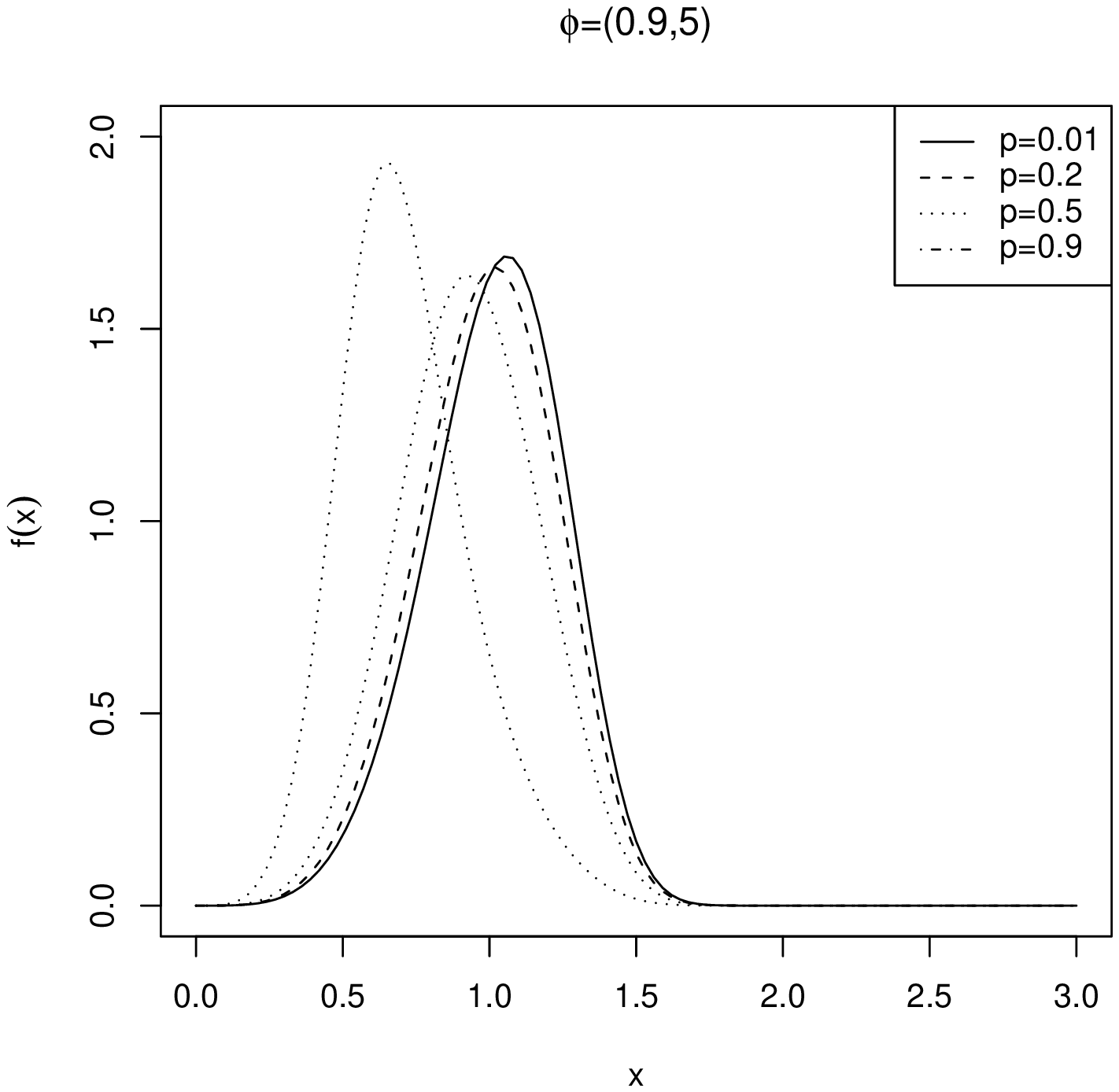}
\caption{Pdf of the WG distribution for selected values of the parameters.}
\label{figura1}
\end{figure}

Suppose that $\{Y_i\}^Z_{i=1}$ are independent and identically
distributed (iid) random variables following the Weibull
distribution $W(\beta,\alpha)$ with scale parameter $\beta>0$, shape
parameter $\alpha >0$ and pdf
\begin{equation*}\label{weibull}
g(x;\beta,\alpha)=\alpha \beta^\alpha x^{\alpha-1} e^{-(\beta
x)^\alpha},\quad x >0,
\end{equation*}
\noindent and $N$ a discrete random variable having a geometric
distribution with probability function $P(n;p)=(1-p)p^{n-1}$ for $n
\in \mathbb{N}$ and $p \in (0,1)$. Let $X=\min{\{Y_i\}^N_{i=1}}$.
The marginal pdf of $X$ is
\begin{equation}\label{WG}
f(x;p,\beta,\alpha)=\alpha \beta^\alpha (1-p) x^{\alpha-1}
e^{-(\beta x)^\alpha}\{1-p \,e^{-(\beta x)^\alpha}\}^{-2},\quad x>0,
\end{equation}
which defines the WG distribution. It is evident that (\ref{WG}) is
much more flexible than the Weibull distribution. The EG
distribution is a special case of the WG distribution for
$\alpha=1$. When $p$ approaches zero, the WG distribution leads to
the Weibull $W(\beta,\alpha)$ distribution. Figure \ref{figura1}
plots the WG density for some values of the vector
$\phi=(\beta,\alpha)$ when $p=0.01,0.2,0.5,0.9$. For all values of
parameters, the density tends to zero as $x\rightarrow\infty$.

For $\alpha>1$, the WG density is unimodal (see appendix A) and the
mode $x_{_0}=\beta^{-1}u_{_0}^{1/\alpha}$ is obtained by solving the
nonlinear equation
\begin{equation}\label{mode}
u_0+p
e^{-u_0}\left(u_0+\frac{\alpha-1}{\alpha}\right)=\frac{\alpha-1}{\alpha}.
\end{equation}
The pdf of the WG distribution can be expressed as an infinite
mixture of Weibull distributions with the same shape parameter
$\alpha$. If $|z|<1$ and $k>0$, we have the series representation
\begin{eqnarray}\label{exp}
(1-z)^{-k}=\sum^{\infty}_{j=0}\frac{\Gamma(k+j)}{\Gamma(k)j!}z^{j}.
\end{eqnarray}

Expanding $\{1-p \,e^{-(\beta x)^\alpha}\}^{-2}$ as in (\ref{exp}),
we can write (\ref{WG}) as
$$f(x;p,\beta,\alpha)=\alpha \beta ^\alpha (1-p) x^{\alpha-1} e^{-(\beta x)^\alpha}\sum^{\infty}_{j=0}(j+1)p^j e^{-j(\beta x)^\alpha}.$$
From the Weibull pdf given before, we have
\begin{equation}\label{combination}
f(x;p,\beta,\alpha)= (1-p) \sum^{\infty}_{j=0} p^j
g(x;\beta(j+1)^{1/\alpha},\alpha).
\end{equation}
Hence, some mathematical properties (cdf, moments, percentiles,
moment generating function, factorial moments, etc.) of the WG
distribution can be obtained using (\ref{combination}) from the
corresponding properties of the Weibull distribution.

\section{Properties of the WG distribution}

\subsection{The distribution and hazard functions and order statistics}

Let $X$ be a random variable such that $X$ follows the WG
distribution with parameters $p$, $\beta$ and $\alpha$. In the
sequel, the distribution of $X$ will be referred to
$WG(p,\beta,\alpha)$. The cdf is given by

\begin{eqnarray}\label{F}
F(x)=\frac{1-e^{-(\beta x)^\alpha}}{1-p\,e^{-(\beta x)^\alpha}}, \quad x>0.
\end{eqnarray}
The survivor and hazard functions are
\begin{equation}\label{survivor}
S(x)=\frac{(1-p)e^{-(\beta x)^\alpha}}{1-p\,e^{-(\beta x)^\alpha}}, \quad x>0 \quad
\end{equation}
\noindent and
\begin{eqnarray}\label{hazard}
h(x)=\alpha \beta^\alpha x^{\alpha-1}\{1-p\,e^{-(\beta x)^\alpha}\}^{-1}, \quad x>0,
\end{eqnarray}
respectively.\\

\begin{figure}[h!]
    \centering
        \includegraphics[width=0.50\textwidth]{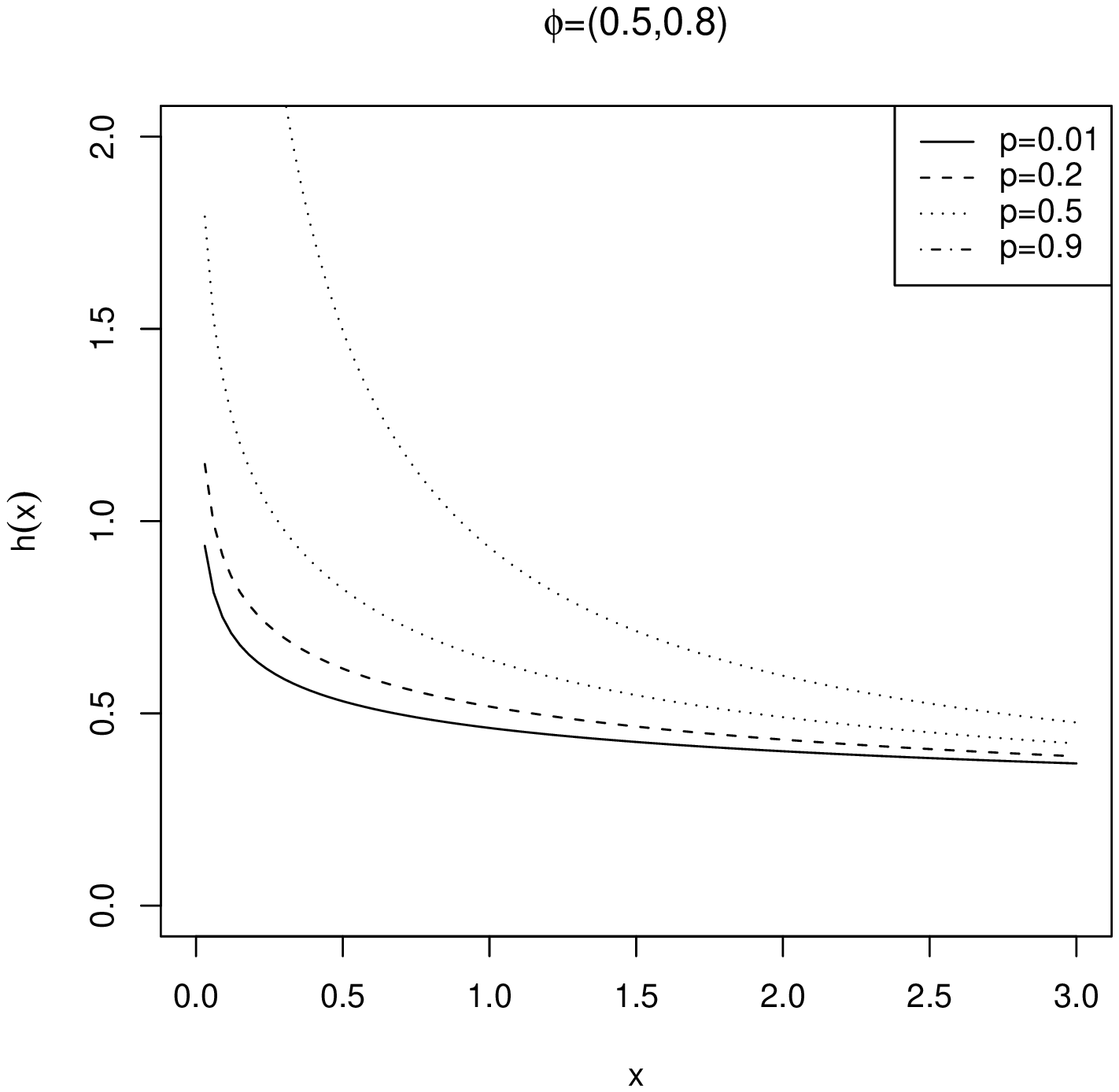}\includegraphics[width=0.5\textwidth]{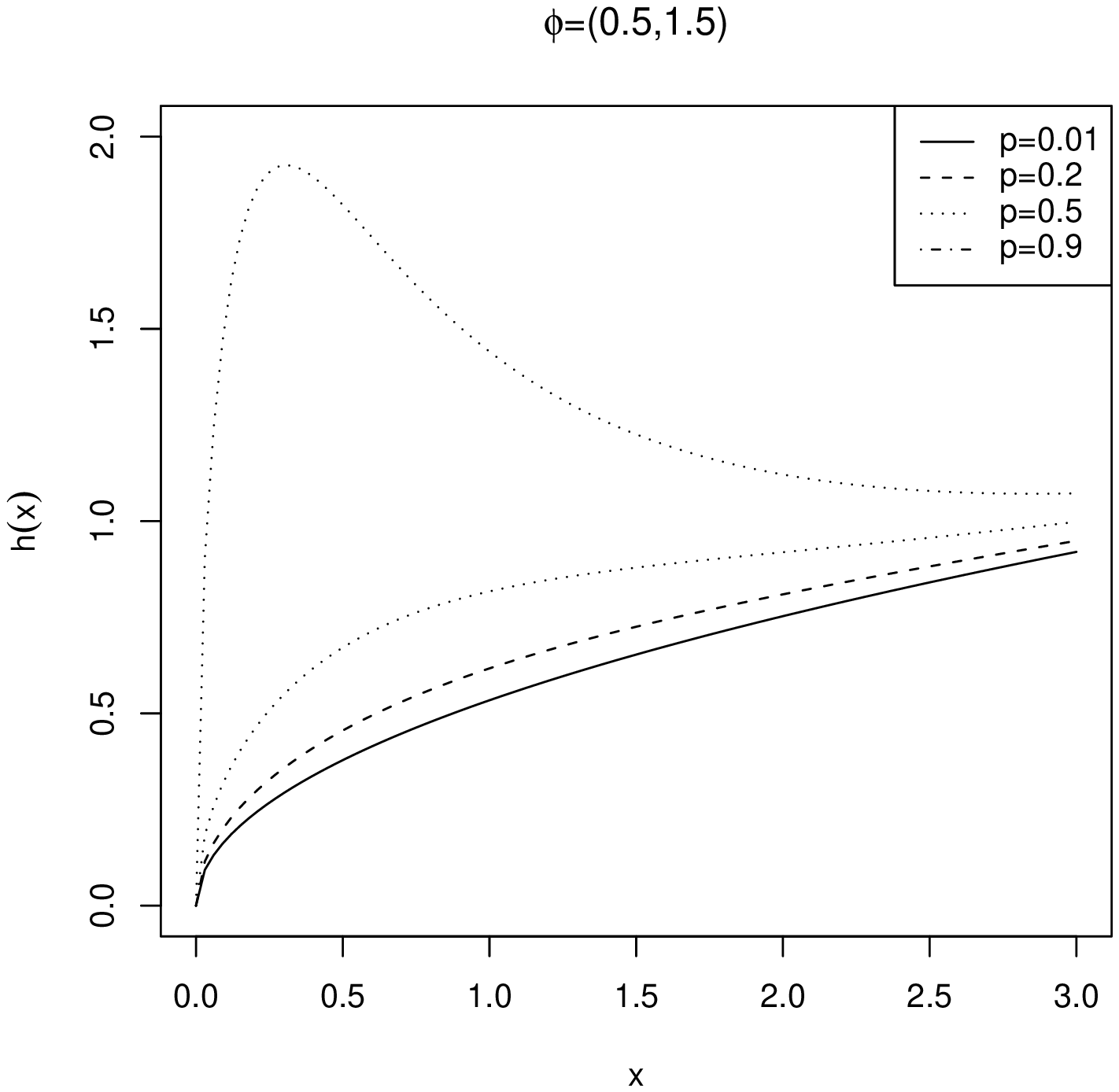}
        \includegraphics[width=0.50\textwidth]{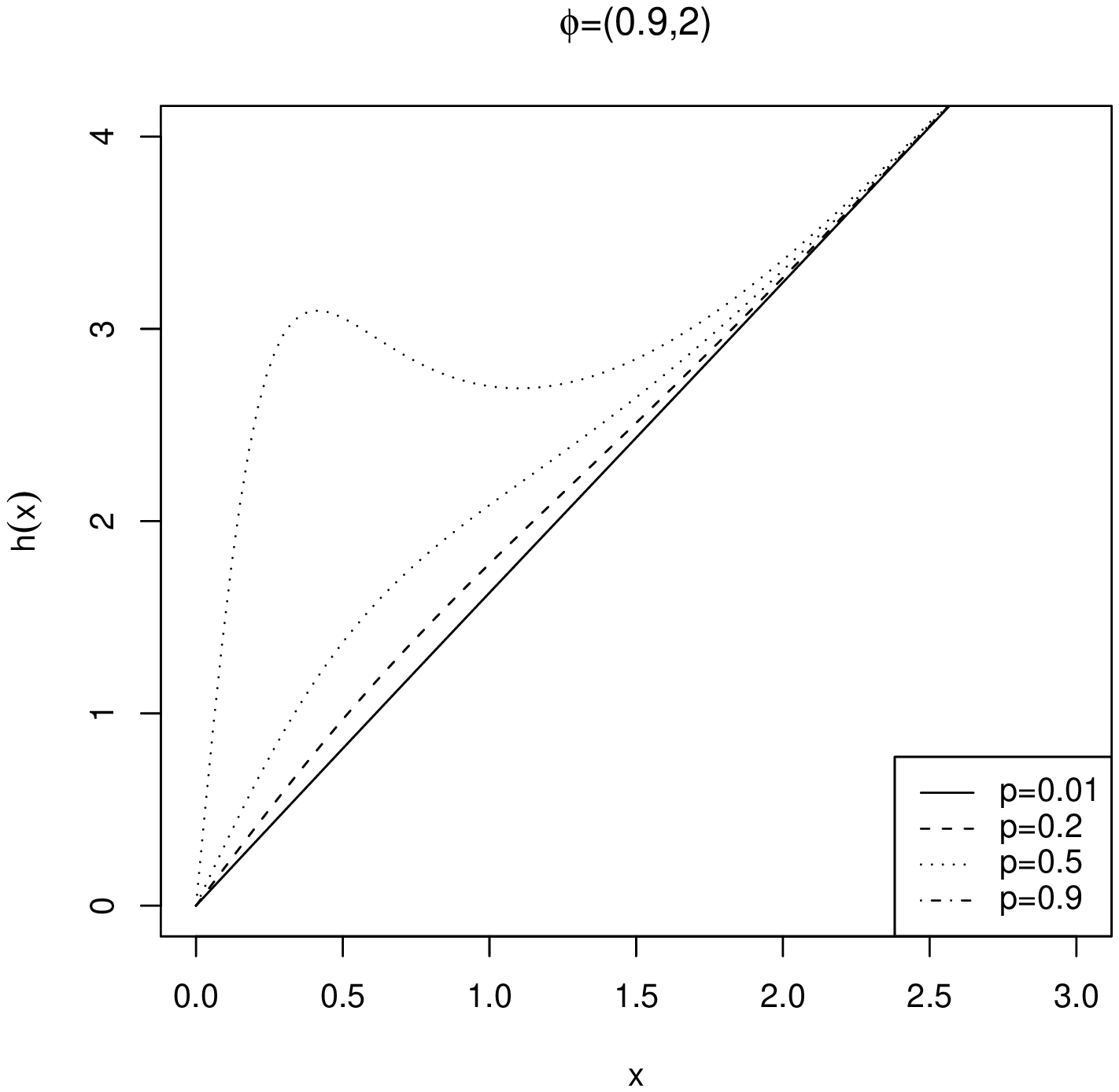}\includegraphics[width=0.5\textwidth]{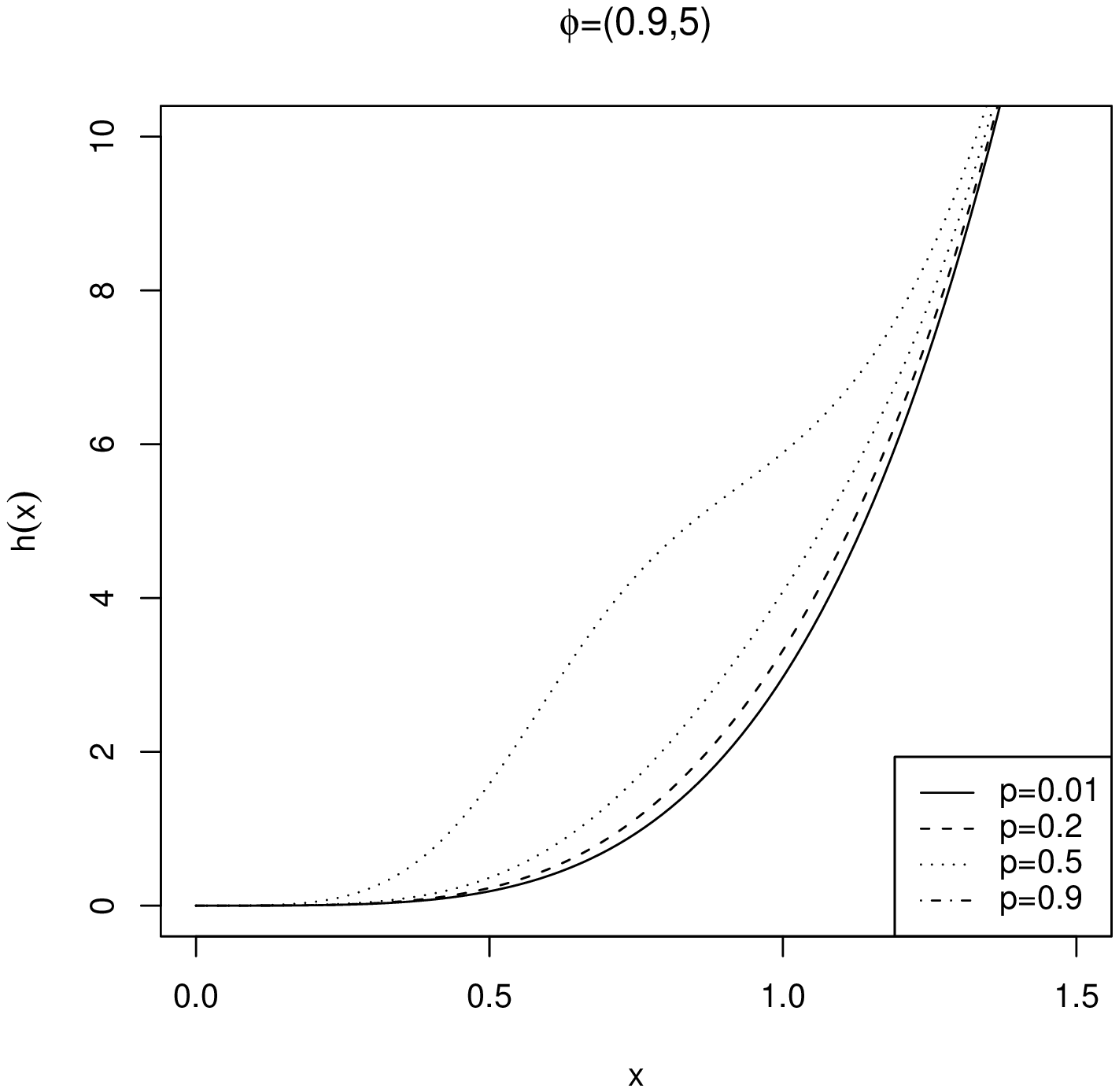}
    \caption{Hazard rate function of the WG distribution for selected values of the parameters.}
    \label{figura2}
\end{figure}

The hazard function (\ref{hazard}) is decreasing for
$0<\alpha\leq1$. However, for $\alpha >1$ it can take different
forms. As the WG distribution converges to the Weibull distribution
when $p\rightarrow 0^+$, the hazard function for very small values
of $p$ can be decreasing, increasing and almost constant. When
$p\rightarrow1^-$, the WG distribution converges to a distribution
degenerate in zero. Hence, the parameter $p$ can be interpreted as a
concentration parameter. Figure \ref{figura2} illustrates some of
the possible shapes of the hazard function for selected values of
the vector $\phi=(\beta,\alpha)$ when $p=0.01,0.2,0.5$ and $0.9$.
These plots show that the hazard function of the new distribution is
quite flexible.\\

We now calculate the pdf of the order statistics. Let
$X_1,\ldots,X_n$ be random variables iid such that $X_i \sim
WG(p,\beta,\alpha)$ for $i=1,\ldots,n$. The pdf of the $i$th order
statistic, $X_{i:n}$ say, is given by (for $x>0$)
\begin{equation}\label{order}
f_{i:n}(x)=\frac{\alpha \beta^\alpha (1-p)^{n-i+1}}{B(i,n-i+1)}x^{\alpha-1}e^{-(n-i+1)(\beta x)^\alpha}\frac{\{1-e^{-(\beta x)^\alpha}\}^{i-1}}{\{1-p e^{-(\beta x)^\alpha}\}^{n+1}},
\end{equation}
\noindent where $B(a,b)=\int_0^\infty\omega^{a-1}(1-\omega)^{b-1} d
\omega$ denotes the beta function. Let $g_{i:n}(x)$ be the pdf of
the $i$th Weibull order statistic with parameters $\beta$ and
$\alpha$ given by
$$g_{i:n}(x)=\frac{\alpha\beta^\alpha}{B(i,n-i+1)}
x^{\alpha-1}e^{-(n-i+1)(\beta x)^\alpha}\{1-e^{-(\beta
x)^\alpha}\}^{i-1}.$$ Equation (\ref{order}) can be rewritten in
terms of $g_{i:n}(x)$ as
\begin{equation*}
f_{i:n}(x)=(1-p)^{n-i+1}\{1-p e^{-(\beta
x)^\alpha}\}^{-(n+1)}g_{i:n}(x).
\end{equation*}
Further, we can express the pdf of $X_{i:n}$ as a mixture of Weibull
order statistic densities. Using (\ref{exp}) in (\ref{order}), we
obtain
\begin{equation}\label{ordercombination}
f_{i:n}(x)=(1-p)^{n-i+1}\frac{n!(n+j-i)!}{(n+j)!(n-i)!}\sum_{j=0}^\infty\binom{n+j}{n}p^j g_{i:n+j}(x).
\end{equation}
Hence, using (\ref{ordercombination}), some mathematical properties
for the order statistics of the WG distribution can be immediately
obtained from the corresponding properties of the Weibull order
statistics.

\subsection{Quantiles and moments}

The quantile $\gamma$ ($x_\gamma$) of the WG distribution follows
from (\ref{F}) as
\begin{eqnarray*}
x_\gamma=\beta^{-1}\left \{\log\left(\frac{1-p\,\gamma}{1-\gamma}\right)\right\}^{1/\alpha}.
\end{eqnarray*}
\noindent In particular, the median is simply
$x_{0.5}=\beta^{-1}\{\log(1-p)\}^{1/\alpha}$.

The $r$th moment of $X$ is given by
\begin{eqnarray*}
E(X^r)=\alpha \beta ^\alpha (1-p)\int_0^\infty x^{r+\alpha-1}
e^{-(\beta x)^\alpha}\left \{ 1 - p \,e^{-(\beta x)^\alpha}\right
\}^{-2}dx.
\end{eqnarray*}
Expanding the term $\{1-p\,e^{-(\beta x)^\alpha}\}^{-2}$ as in
(\ref{exp}) yields
\begin{eqnarray*}
E(X^r)=\frac{(1-p)\Gamma(r/\alpha+1)}{p \,\beta^r }L(p;r/\alpha),
\end{eqnarray*}
where $L(p;a)=\sum_{j=1}^{\infty}p^j j^{-a}$ is Euler's
polylogarithm function (see, Erdelyi et al., 1953, p. 31) which is
readily available in standard software such as Mathematica.\\

\begin{figure}[h!]
\centering
\includegraphics[width=0.5 \textwidth]{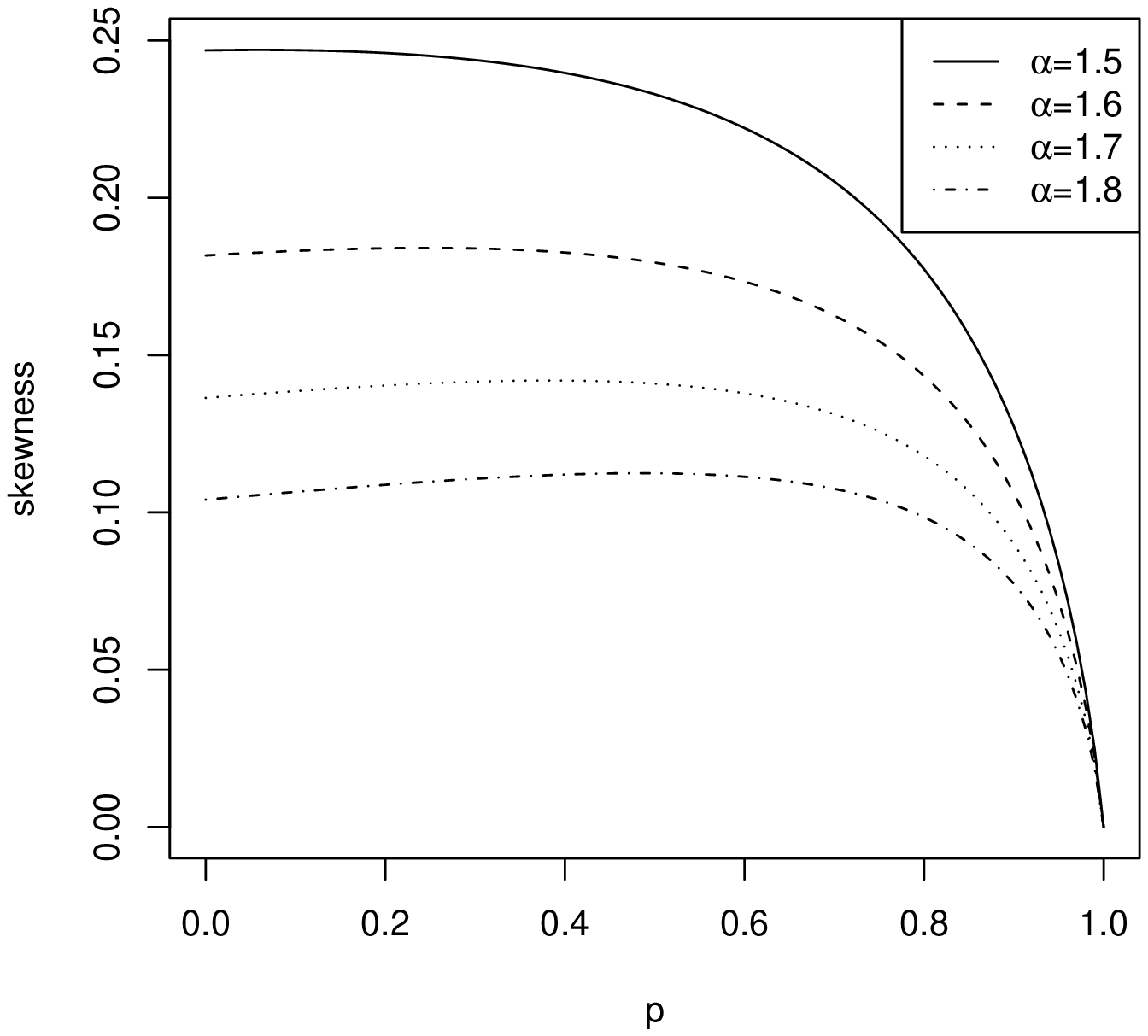}\includegraphics[width=0.5 \textwidth]{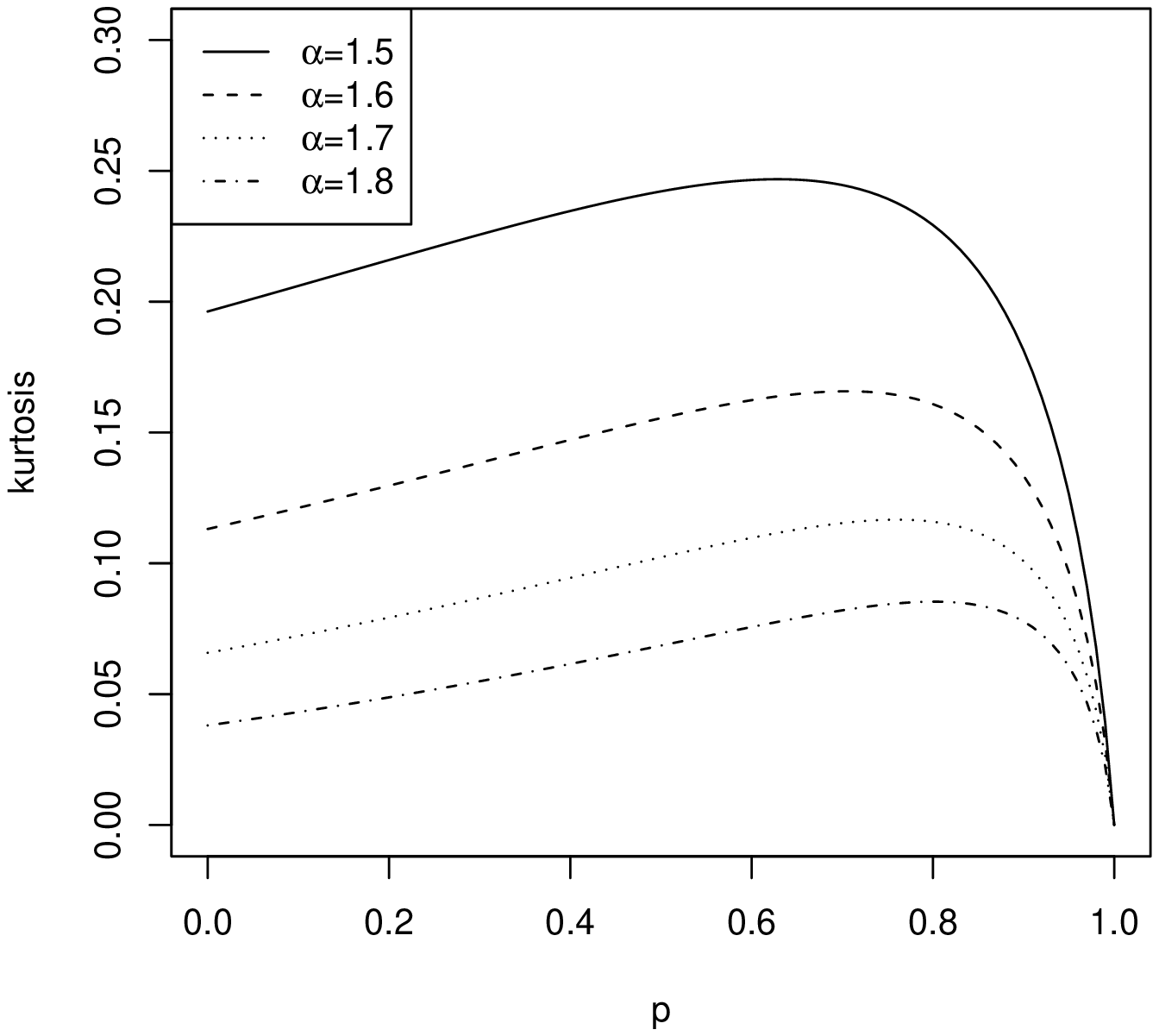}
\caption{Skewness and kurtosis of the WG distribution as functions
of $p$ for $\beta=1$ and some values of $\alpha$.} \label{skew}
\end{figure}

Figure \ref{skew} plots the skewness and kurtosis of the WG
distribution as functions of $p$ for $\beta=1$ and some values of
$\alpha$. When $p\rightarrow1^-$, the coefficients of skewness and
kurtosis tend to zero as expected, since the WG distribution
converges to a degenerate distribution (in zero) when
$p\rightarrow1^-$.

The $r$th moment of the $i$th order statistic $X_{i:n}$ is given by
\begin{eqnarray*}
E(X^r_{i:n})=\frac{\alpha \beta^\alpha (1-p)^{n-i+1}}{B(i,n-i+1)}\int_0^\infty x^{\alpha+r-1} e^{-(n-i+1)(\beta x)^\alpha}\frac{\{1-e^{-(\beta x)^\alpha}\}^{i-1}}{\{1-p e^{-(\beta x)^\alpha}\}^{n+1}}dx.
\end{eqnarray*}
Expanding the term $\{1-pe^{-(\beta x)^\alpha}\}^{-(n+1)}$ as in
(\ref{exp}) and using the binomial expansion for $\{1-e^{-(\beta
x)^\alpha}\}^{i-1}$, the $r$th moment of $X_{i:n}$ becomes
\begin{equation}\label{momorder}
E(X_{i:n}^r)=\frac{(1-p)^{n-i+1}\Gamma(r/\alpha+1)}
{B(i,n-i+1)\beta^r}\sum_{j=0}^\infty \sum_{k=0}^{i-1}\frac{(-1)^k \binom{n+j}{n}\binom{i-1}{k}p^j}{(n+j+k-i+1)^{r/\alpha+1}}.
\end{equation}
We now give an alternative expression to (\ref{momorder}) by using a
result due to Barakat and Abdelkader (2004). We have
\begin{eqnarray*}
E(X_{i:n}^r)=r\sum_{k=n-i+1}^n (-1)^{k-n+i-1}\binom{k-1}{n-i}\binom{n}{k}\int_0^\infty x^{r-1}S(x)^kdx,
\end{eqnarray*}
where $S(x)$ is the survivor function (\ref{survivor}).\\

Using the expansion (\ref{exp}) and changing variables
$u=(k+j)(\beta x)^\alpha$, we have
\begin{eqnarray*}
\int_0^\infty x^{r-1}S(x)^kdx&=&(1-p)^k\sum_{j=0}^\infty\binom{k+j-1}{k-1}p^j\int_0^\infty x^{r-1}e^{-(k+j)(\beta x)^\alpha}dx\\
&=&\frac{(1-p)^k}{\alpha\beta^r}\int_0^\infty u^{r/\alpha-1}e^{-u}du\sum_{j=0}^\infty\binom{k+j-1}{k-1}\frac{p^j}{(k+j)^{r/\alpha}}\\
&=& \frac{(1-p)^k\Gamma(r/\alpha)}{\alpha\beta^r}\sum_{j=0}^\infty \binom{k+j-1}{k-1}\frac{p^j}{(k+j)^{r/\alpha}}.
\end{eqnarray*}
Hence,
\begin{eqnarray}\label{alt}
E(X_{i:n}^r)&=&\frac{\Gamma(r/\alpha+1)}{(-1)^{n-i+1}\beta^r}\sum_{j=0}^\infty\sum_{k=n-i+1}^n(-1)^{k}\binom{n}{k}\binom{k-1}{n-i}
\binom{k+j-1}{k-1}\frac{p^j(1-p)^k}{(k+j)^{r/\alpha}}.\nonumber\\
\end{eqnarray}
Expressions (\ref{momorder}) and (\ref{alt}) give the moments of the
order statistics and can be compared numerically. Table \ref{t1}
gives numerical values for the first four moments of the order
statistics $X_{1:15}$, $X_{7:15}$ and $X_{15:15}$ from
(\ref{momorder}) and (\ref{alt}) with the index $j$ stopping at
$100$ and by numerical integration. We take the parameter values as
$p=0.8$, $\beta=0.4$ and $\alpha=2$. The figures in this table show
good agreement among the three methods.

\begin{table}[!h]
\begin{center}
\begin{tabular}{cc|cccc}
\hline
$X_{i:15}\downarrow$&$r$th moment$\rightarrow$&$r=1$&$r=2$&$r=3$&$r=4$\\
\hline
&Expression (\ref{momorder})&0.25717&0.08697&0.035116&0.016265\\
$i=1$&Expression (\ref{alt})&0.25717&0.08697&0.035116&0.016265\\
&Numerical&0.26102&0.08795&0.035408&0.016364\\
\hline
&Expression (\ref{momorder})&0.96660&0.98827&1.06643&1.21249\\
$i=7$&Expression (\ref{alt})&0.98502&0.99295&1.06784&1.21298\\
&Numerical&0.96674&0.98836&1.06649&1.21253\\
\hline
&Expression (\ref{momorder})&3.33109&11.97872&46.35371&192.32090\\
$i=15$&Expression (\ref{alt})&3.33126&11.97875&46.35375&192.32090\\
&Numerical&3.33126&11.97875&46.35375&192.32090\\
\hline
\end{tabular}\caption{First four moments of some order statistics from (\ref{momorder}) and (\ref{alt}) and by numerical integration.}\label{t1}
\end{center}
\end{table}

\subsection{Rényi and Shannon entropies}

Entropy has been used in various situations in science and
engineering. The entropy of a random variable $X$ is a measure of
variation of the uncertainty. Rényi entropy is defined by
$I_R(\gamma)=\frac{1}{1-\gamma}\log\{\int_{\mathbb{R}}
f^\gamma(x)dx\},$ where $\gamma>0$ and $\gamma\neq1$. By using
(\ref{exp}), we have
\begin{eqnarray*}
\int_0^\infty f^\gamma(x;p,\beta,\alpha)dx=\frac{[\alpha\beta^\alpha(1-p)]^\gamma}{\Gamma(2\gamma)}\sum_{j=0}^\infty p^j\frac{\Gamma(2\gamma+j)}{j!}\int_0^\infty x^{(\alpha-1)\gamma}e^{-(\gamma+j)(\beta x)^\alpha}dx.
\end{eqnarray*}
If $(\alpha-1)(\gamma-1)\geq0$, this expression reduces to
\begin{eqnarray*}
\int_0^\infty f^\gamma(x;p,\beta,\alpha)dx=\frac{\Gamma(\alpha)[\alpha(1-p)]^\gamma}{\beta^{\alpha(1-\gamma)}\Gamma(2\gamma)}\sum_{j=0}^\infty p^j\frac{\Gamma(2\gamma+j)}{j!(\alpha+j)}E(Y_j^{(\alpha-1)(\gamma-1)}),
\end{eqnarray*}
where $Y_j$ follows a gamma distribution with scale parameter
$(\gamma+j)^{1/\alpha}$ and shape parameter $\alpha$. Hence, we
obtain
\begin{eqnarray*}
I_R(\gamma)=\frac{1}{1-\gamma}\log\left\{\frac{[\alpha(1-p)]^\gamma\Gamma(\gamma(\alpha-1)+1)}{\beta^{1-\gamma}\Gamma(2\gamma)}\sum_{j=0}^\infty \frac{p^j\Gamma(2\gamma+j)}{j!(\alpha+j)^{(\alpha-1)(\gamma-1)/\alpha+1}}\right\}.
\end{eqnarray*}

Shannon entropy is defined as $E\{-\log[f(X)]\}$. This is a special
case obtained from $\lim_{\gamma\rightarrow1}I_R(\gamma)$. Then,
$$E[-\log f(X)]=-\log[\alpha\beta^\alpha(1-p)]-(\alpha-1)E[\log(X)]+\beta^\alpha E(X^\alpha)-2 E\{\log[1-pe^{-(\beta X)^\alpha}]\}.$$
We can show that
\begin{eqnarray*}
E[\log(X)]&=&\psi(1)/\alpha, \\
E(X^\alpha)&=&-\frac{(1-p)}{p\beta^\alpha}\log(1-p),\\
E\{\log[1-pe^{-(\beta X)^\alpha}]\}&=&-\frac{1-p}{p}\{1+(1-p)[1+\log(1-p)]\}.
\end{eqnarray*}
Hence, the Shannon entropy reduces to
\begin{eqnarray*}
E[-\log
f(X)]=-\log[\alpha\beta^\alpha(1-p)]-\frac{\alpha-1}{\alpha}\psi(1)-\frac{1-p}{p}[4-2p+(3-2p)\log(1-p)].
\end{eqnarray*}

\section{Estimation}

Let $x=(x_1,\ldots,x_n)$ be a random sample of the WG distribution
with unknown parameter vector $\theta=(p,\beta,\alpha)$. The log
likelihood $\ell=\ell(\theta;x)$ for $\theta$ is
\begin{eqnarray*}
\ell&=&n[\log\alpha+\alpha \log \beta+\log(1-p)]+
(\alpha-1)\sum^n_{i=1}\log(x_i)-\sum^n_{i=1}(\beta x_i)^\alpha\\
&-&2\sum^n_{i=1}\log[1-p\,e^{-(\beta x_i)^\alpha}].
\end{eqnarray*}
The score function $U(\theta)=(\partial \ell/\partial p,\partial
\ell/\partial \beta,\partial \ell/\partial \alpha)^T$ has components
\begin{eqnarray*}
\frac{\partial \ell}{\partial p}&=&-n(1-p)^{-1}+2\sum^n_{i=1}e^{-(\beta x_i)^\alpha}[1-p\,e^{-(\beta x_i)^\alpha}]^{-1} ,\\
\frac{\partial \ell}{\partial \beta}&=&n\alpha\beta^{-1}-\alpha
\beta^{\alpha -1 }\sum^n_{i=1} x_i^\alpha \{1+ 2p\,e^{-(\beta
x_i)^\alpha}[1-p\,e^{-(\beta x_i)^\alpha}]^{-1}\},\\
\frac{\partial \ell}{\partial \alpha}&=&n\alpha^{-1}+\sum^n_{i=1}\log(\beta x_i)-\sum^n_{i=1}(\beta x_i)^\alpha \log(\beta x_i)\{1+ 2p\,e^{-(\beta x_i)^\alpha}[1-p\,e^{-(\beta x_i)^\alpha}]^{-1}\}.\\
\end{eqnarray*}

The maximum likelihood estimate (MLE) $\widehat\theta$ of $\theta$
is calculated numerically from the nonlinear equations
$U(\theta)=0$. We use the EM algorithm (Dempster et al., 1977;
McLachlan and Krishnan, 1997) to obtain $\widehat\theta$. For doing
this, we define an hypothetical complete-data distribution with
density function
$$f(x,z;\theta)=\alpha \beta^\alpha(1-p)z
x^{\alpha-1}p^{z-1}e^{-z(\beta x)^\alpha},$$ for $x,\beta,\alpha>0$,
$p \in (0,1)$ and $z \in \mathbb{N}$. Under this formulation, the
E-step of an EM cycle requires the conditional expectation of
$(Z|X;\theta^{(r)})$, where
$\theta^{(r)}=(p^{(r)},\beta^{(r)},\alpha^{(r)})$ is the current
estimate of $\theta$. Using that $P(z|x;\theta)=z p^{z-1}
e^{-(z-1)(\beta x)^\alpha}\times$ $\{1-pe^{-(\beta x)^\alpha}\}^2$
for $z \in N$, it follows $E(Z|X;\theta)=\{1 + p e^{-(\beta
x)^\alpha}\}\{1-pe^{-(\beta x)^\alpha}\}^{-1}$. The EM cycle is
completed with the M-step by using the maximum likelihood estimation
over $\theta$, with the missing $Z$'s replaced by their conditional
expectations given above. Hence, an EM iteration reduces to
$$p^{(r+1)}=1-\frac{n}{\sum^n_{i=1}w_i^{(r)}},\,\,\,\,
\beta^{(r+1)}=n\left\{\sum^n_{i=1}x_i^{\alpha^{(r+1)}}w_i^{(r)}\right\}^{-1/\alpha^{(r+1)}},$$
where $\alpha^{(r+1)}$ is the solution of the nonlinear equation
$$\frac{n}{\alpha^{(r+1)}}+\sum^n_{i=1}\log x_i-n\frac{\sum^n_{i=1}w_i^{(r)} x_i^{\alpha^{(r+1)}}\log
x_i}{\sum^n_{i=1}w_i^{(r)} x_i^{\alpha^{(r+1)}}}=0,$$
where
\begin{equation*}
w_i^{(r)}=\frac{1+p^{(r)}e^{-(\beta^{(r)} x_i)^{\alpha^{(r)}}}}{1-p^{(r)}e^{-(\beta^{(r)} x_i)^{\alpha^{(r)}}}}.
\end{equation*}
An implementation of this algorithm using the software R is given in
Appendix B.

\section{Inference}

For interval estimation and hypothesis tests on the model
parameters, we require the information matrix. The $3\times3$
observed information matrix $J_n=J_n(\theta)$ is given by
$$J_n = \left(\begin{array}{cccc}
J_{pp}&J_{p\beta}&J_{p\alpha}\\
J_{p\beta}&J_{\beta\beta}&J_{\beta\alpha}\\
J_{p\alpha}&J_{\beta\alpha}&J_{\alpha\alpha}\\
\end{array}\right),$$
where

\begin{eqnarray*}
-J_{pp}&=&\frac{\partial^2\ell}{\partial p^2}=2 \sum_{i=1}^nT^{(i)}_{0,0,2,2}-n(1-p)^{-2},\\
-J_{p\alpha}&=&\frac{\partial^2\ell}{\partial p \partial \alpha}=-2\beta^{\alpha}\sum_{i=1}^n( p\, T^{(i)}_{1,1,2,2}+ T^{(i)}_{1,1,1,1}),\\
-J_{p\beta}&=&\frac{\partial^2\ell}{\partial p \partial \beta }= -2 \alpha \beta^{\alpha-1}\sum_{i=1}^n(p\,T^{(i)}_{1,0,2,2}+T^{(i)}_{1,0,1,1}),\\
-J_{\alpha\alpha}&=&\frac{\partial^2\ell}{\partial \alpha^2}=-n\alpha^{-2}+\sum_{i=1}^n(2 p^2 \beta^{2\alpha} T^{(i)}_{2,2,2,2}+2 p \beta^{2\alpha}T^{(i)}_{2,2,1,1}-\beta^\alpha T^{(i)}_{1,2,0,0}-\\
&&2p\beta^\alpha T^{(i)}_{1,2,1,1}),\\
-J_{\beta\alpha}&=&\frac{\partial^2\ell}{\partial \alpha \partial \beta}= n\beta^{-1}-\beta^{\alpha -1}\sum_{i=1}^n (\alpha T^{(i)}_{1,1,0,0}+T^{(i)}_{1,0,0,0})(1+2 p T^{(i)}_{0,0,1,1})+\\
&&2p\alpha \beta^{2\alpha -1}\sum_{i=1}^n (p\,T^{(i)}_{2,1,2,2}+T^{(i)}_{2,1,1,1}),
\end{eqnarray*}
\begin{eqnarray*}
-J_{\beta\beta}&=&\frac{\partial^2\ell}{\partial \beta^2}=-n\alpha \beta^{-2}-\alpha(\alpha-1)\beta^{\alpha-2}\sum_{i=1}^n(T^{(i)}_{1,0,0,0}+2 pT^{(i)}_{1,0,1,1})+\\
&&2\alpha^2\beta^{2\alpha-2}p\sum_{i=1}^n(p T^{(i)}_{2,0,2,2}+T^{(i)}_{2,0,1,1}).\\
\end{eqnarray*}
Here,
\begin{eqnarray*}
T_{j,k,l,m}^{(i)}=T_{j,k,l,m}^{(i)}(x_i,\theta)= x_i^{\alpha j}
\{\log(\beta x_i)\}^k\{1-p\,e^{-(\beta x_i)^\alpha}\}^{-l}
e^{-m(\beta x_i)^\alpha},
\end{eqnarray*}
for $(j,k,l,m)\in\{0,1,2\}$ and $i=1,\ldots,n$. Under conditions
that are fulfilled for the parameter $\theta$ in the interior of the
parameter space but not on the boundary, the asymptotic distribution
of $\sqrt n
(\widehat\theta-\theta)\,\,\,\,\mathrm{is\,multivariate\,
normal}\,\,\,\,N_3(0,K(\theta)^{-1})$, where
$K(\theta)=\lim_{n\rightarrow\infty} n^{-1} J_n(\theta)$ is the unit
information matrix. This asymptotic behavior remains valid if
$K(\theta)$ is replaced by the average sample information matrix
evaluated at $\widehat\theta$, i.e., $n^{-1}J_n(\widehat\theta)$. We
can use the asymptotic multivariate normal
$N_3(0,J_n(\widehat\theta)^{-1})$ distribution of $\widehat\theta$
to cons\-truct approximate confidence regions for some parameters
and for the hazard and survival functions. In fact, an
$100(1-\gamma)\%$ asymptotic confidence interval for each parameter
$\theta_i$ is given by
$$ACI_i=(\widehat\theta_i-z_{\gamma/2}\sqrt{\widehat J^{\theta_i\theta_i}},\widehat{\theta_i}
+z_{\gamma/2}\sqrt{\widehat J^{\theta_i\theta_i}}),$$
where $\widehat J^{\theta_i\theta_i}$ represents the $(i,i)$ diagonal element of
$J_n(\widehat\theta)^{-1}$ for $i=1,2,3,4$ and $z_{\gamma/2}$ is the quantile $1-\gamma/2$
of the standard normal distribution.

The asymptotic normality is also useful for testing goodness of fit
of the three parameter WG distribution and for comparing this
distribution with some of its special sub-models via the likelihood
ratio (LR) statistic. We consider the partition
$\theta=(\theta_1^T,\theta_2^T)^T$, where $\theta_1$ is a subset of
parameters of interest of the WG distribution and $\theta_2$ is a
subset of the remaining parameters. The LR statistic for testing the
null hypothesis $H_0:\theta_1 =\theta_1^{(0)}$ versus the
alternative hypothesis $H_1:\theta_1 \neq \theta_1^{(0)}$ is given
by $w= 2\{\ell(\hat{\theta})-\ell(\tilde{\theta})\}$, where
$\tilde\theta$ and $\widehat\theta$ denote the MLEs under the null
and alternative hypotheses, respectively. The statistic $w$ is
asymptotically (as $n\to\infty$) distributed as $\chi_k^2$, where
$k$ is the dimension of the subset $\theta_1$ of interest. For
example, we can compare the EG model against the WG model by testing
$H_0:\alpha=1$ versus $H_1:\alpha \ne 1$ and the Weibull model
against the WG model by testing $H_0:\alpha=1,p=0$ versus $H_1:{\rm
\,\,H_0\,\,is\,\,false}$.

\section{Applications}

In this section, we fit the WG models to two real data sets. The
first data set consist of the number of successive failures for the
air conditioning system of each member in a fleet of 13 Boeing 720
jet airplanes. The pooled data with 214 observations were first
analyzed by Proschan (1963) and discussed further by Dahiya and
Gurland (1972), Gleser (1989), Adamidis and Loukas (1998) and Kus
(2007). The second data set is an uncensored data set from Nichols and Padgett (2006) consisting of
100 observations on breaking stress of carbon fibres (in Gba).\\

\begin{figure}[h!]
\centering
\includegraphics[width=0.5 \textwidth]{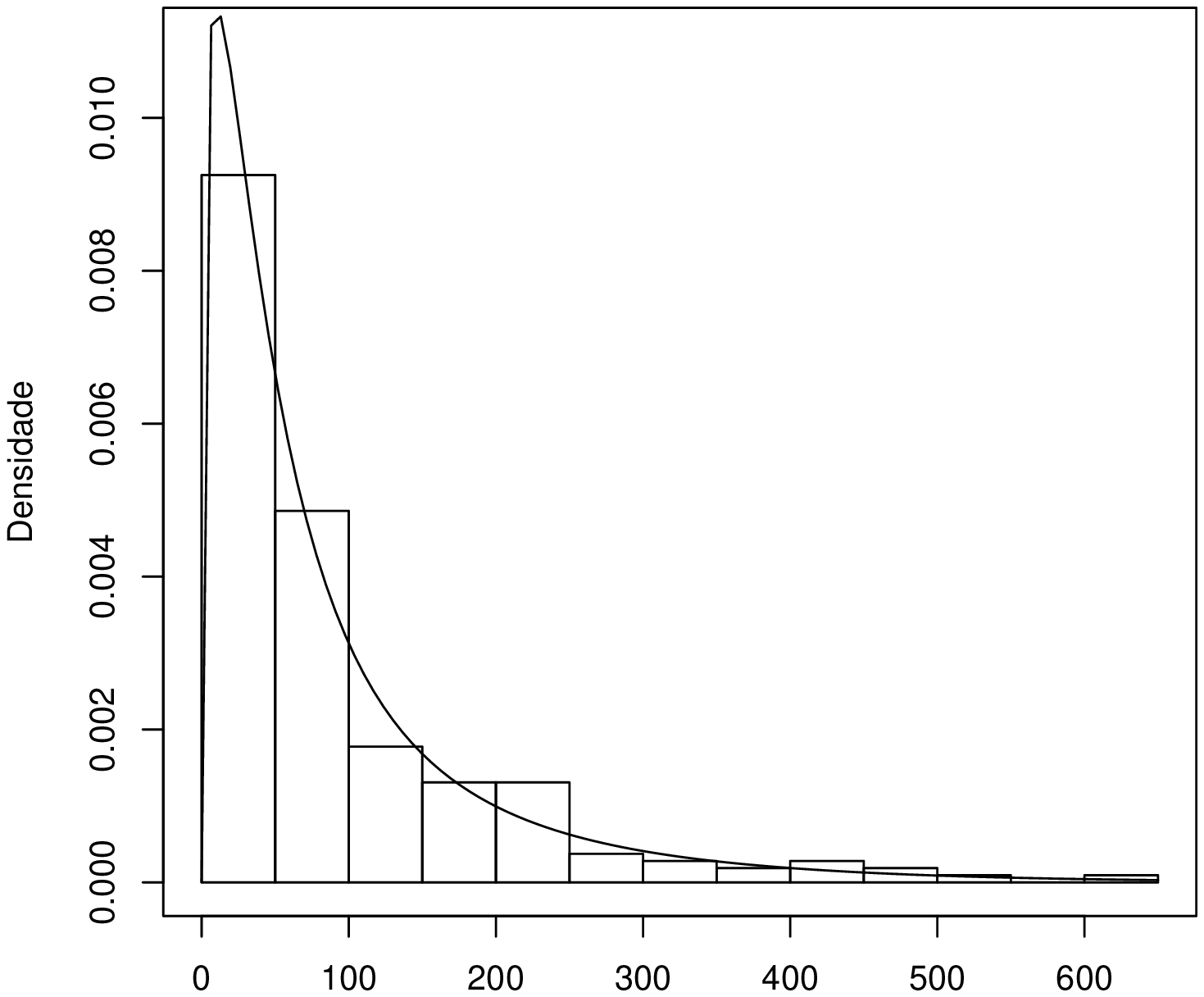}\includegraphics[width=0.5 \textwidth]{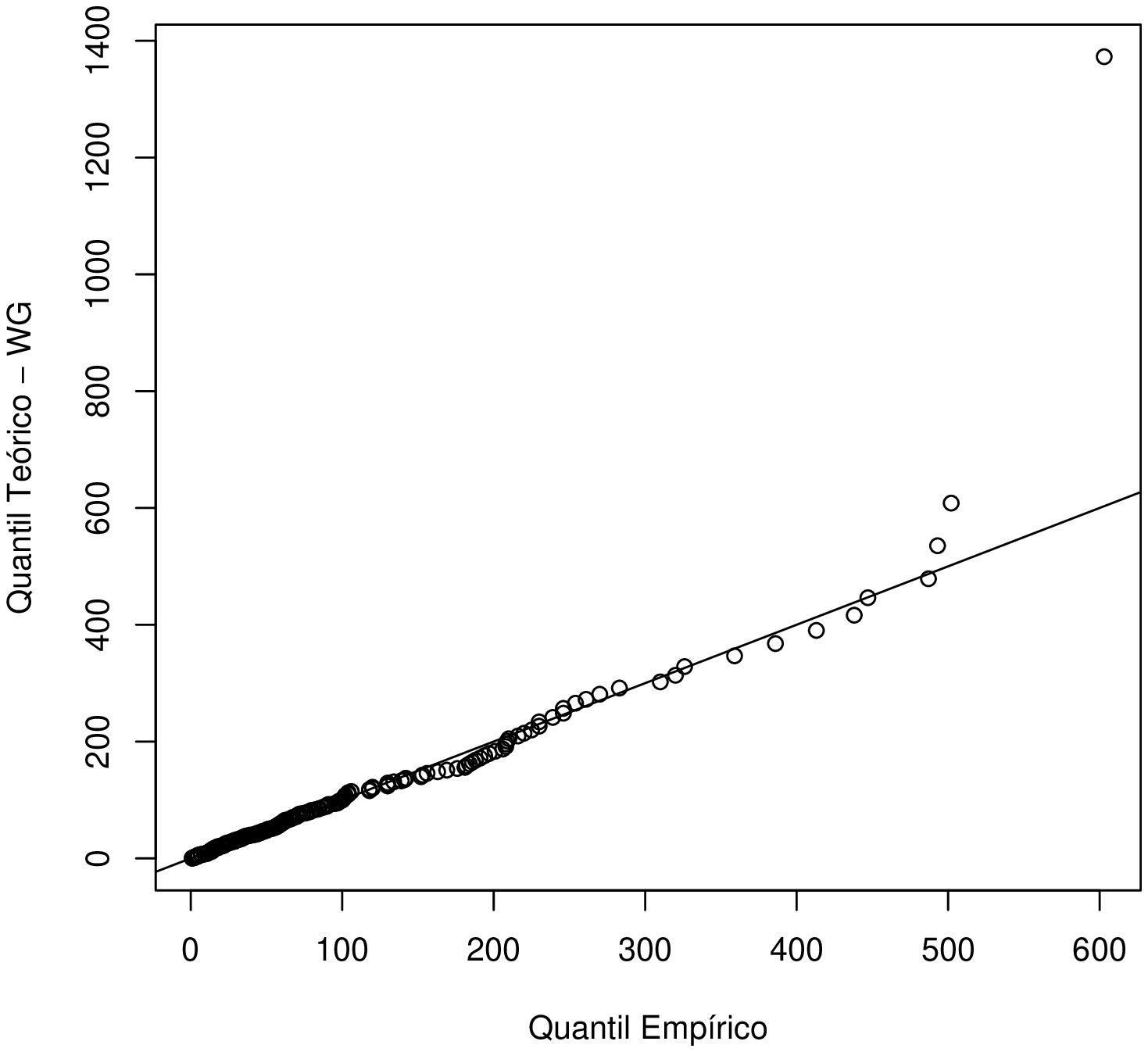}
\caption{Plots of the fitted pdf and of the estimated quantiles
versus observed quantiles for the first data set.}
\label{densidade1}
\end{figure}

For the first data set, the estimated parameters using an EM
algorithm were $\hat{p}=0.7841$, $\hat{\beta}=0.0048$ and
$\hat{\alpha}=1.2246$. The fitted pdf and the estimated quantiles
versus observed quantiles are given in Figures \ref{densidade1}.
This figure shows a good fit of the WG model for the first data set.

\begin{figure}[h!]
\centering
\includegraphics[width=0.5 \textwidth]{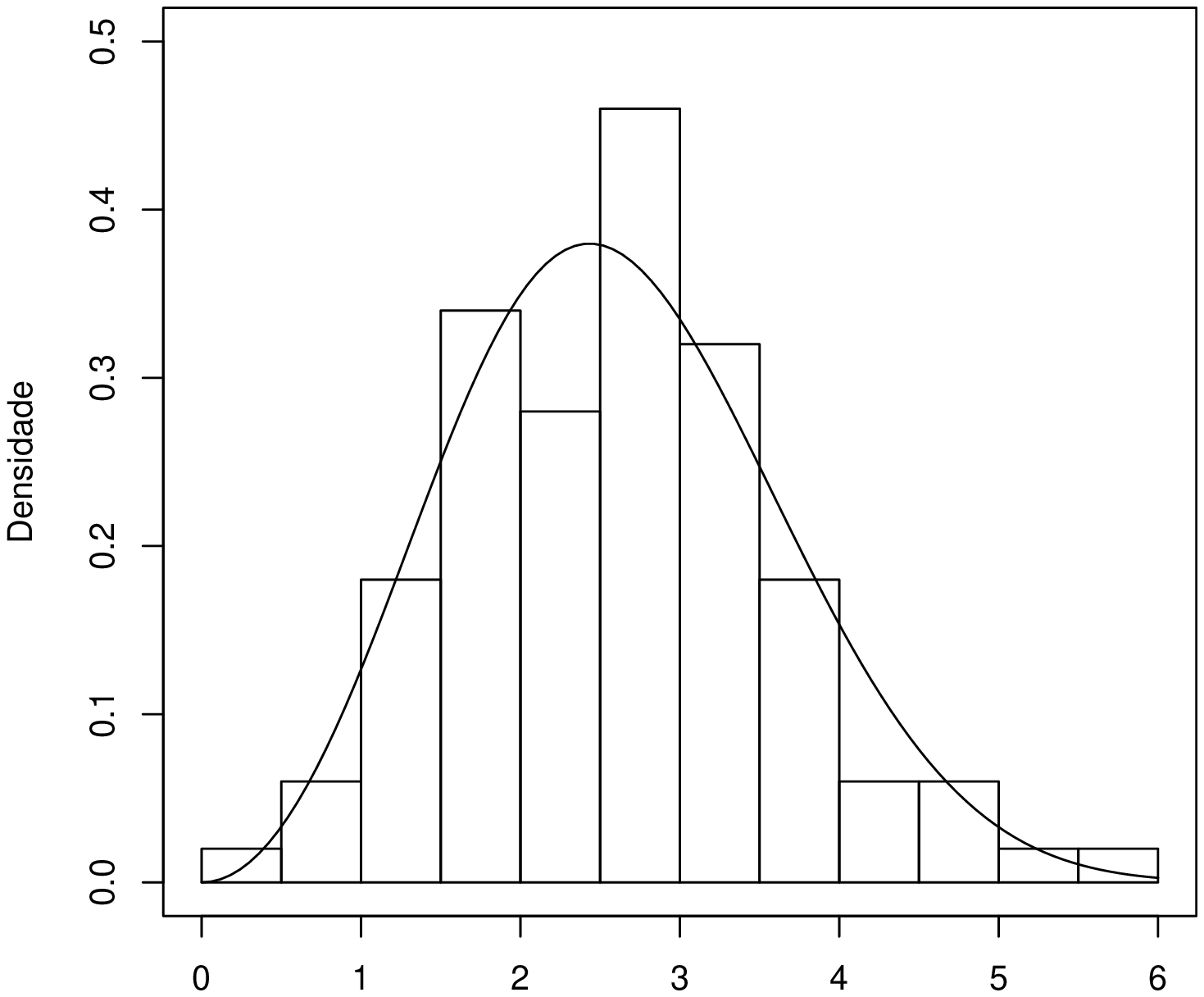}\includegraphics[width=0.5 \textwidth]{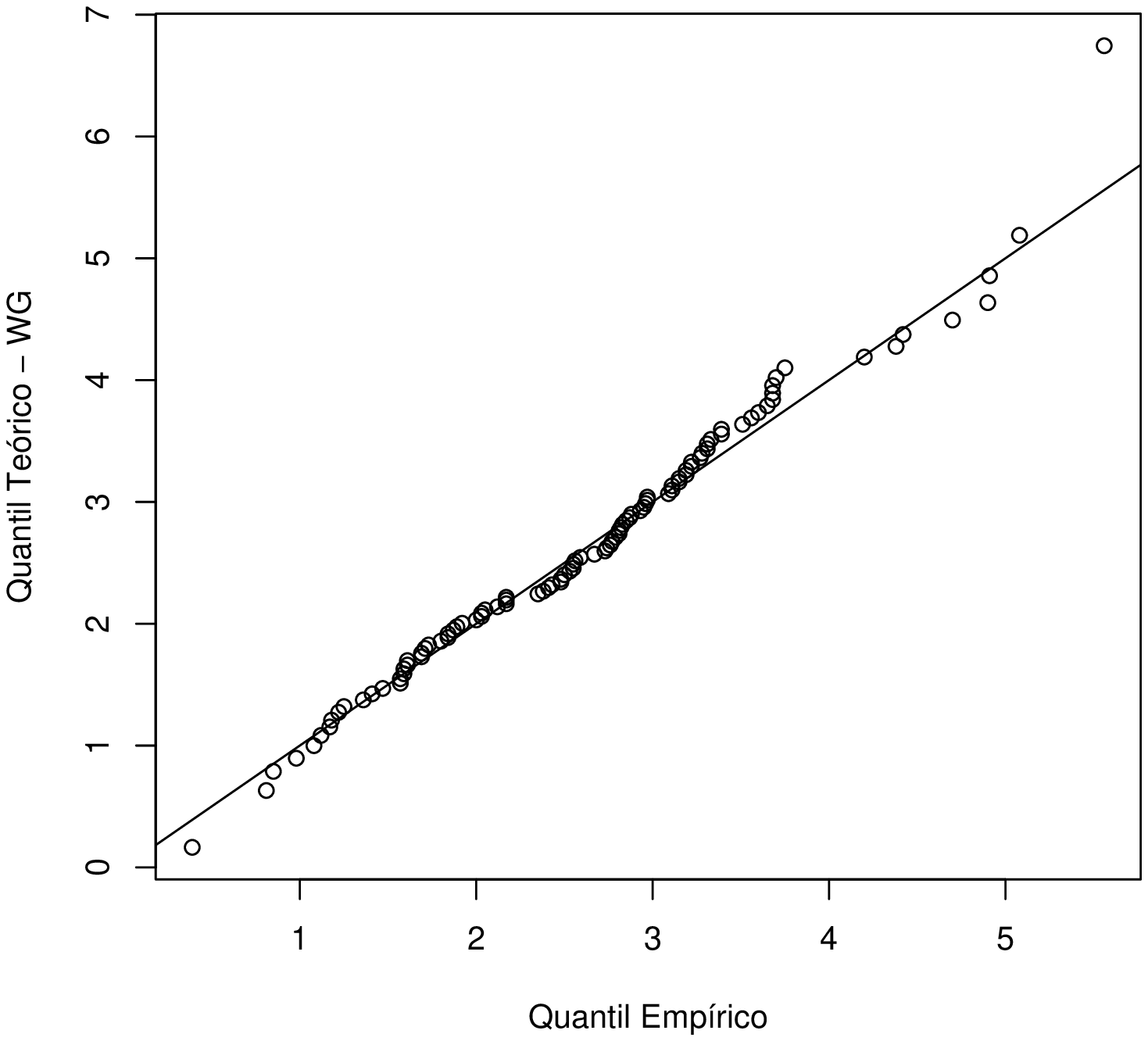}
\caption{Plots of the fitted pdf and of the estimated quantiles
versus observed quantiles for the second data set.}
\label{densidade}
\end{figure}

For the second data set, the estimates obtained using an EM
algorithm are $\hat{p}=0.3073$, $\hat{\beta}= 0.3148$ and
$\hat{\alpha}= 3.0093$. The plot of the fitted pdf and the estimated
quantiles versus observed quantiles in Figure \ref{densidade} shows
a good fit of the WG model.

\section{Conclusions}

We define a new model so called the Weibull-geometric (WG)
distribution that generalizes the exponential-geometric (EG)
distribution introduced by Adamidis and Loukas (1998). Some
mathematical properties are derived and plots of the pdf and hazard
functions are presented to show the flexibility of the new
distribution. We give closed form expressions for the moments of the
distribution. We obtain the pdf of the order statistics and provide
expansions for the moments of the order statistics. Estimation by
maximum likelihood is discussed and an algorithm EM is proposed. We
discuss inference, give asymptotic confidence intervals for the
model parameters and present the use of the LR statistic to compare
the fit of the WG model with special sub-models. Finally, we fitted
WG models to two real data sets to show the flexibility and the
potentially of the new distribution.


\section*{Appendix A}

We now show that the WG density is unimodal when $\alpha>1$. Let
$$h(u)=u+p e^{-u}\left(u+\frac{\alpha-1}{\alpha}\right).$$

When $u \rightarrow 0^+$, $h(u)\rightarrow p\frac{\alpha-1}{\alpha}$
and when $u \rightarrow \infty$, $h(u)\rightarrow \infty$. Thus, if
$h(u)$ is an increasing function, the solution in (\ref{mode}) is
unique and the WG distribution is unimodal. We have $h'(u)=1-p
e^{-u}\left(u+\frac{\alpha-1}{\alpha}\right)+p e^{-u}$. Using the
inequalities $-p e^{-u}\frac{\alpha-1}{\alpha}>-p e^{-u}$ and
$-pe^{-u}u> e{-1}$, it follows that $h'(u)>1-e^{-1}>0$, $\forall
u>0$. Hence, $h(u)$ is an increasing function and the WG
distribution is unimodal if $\alpha>1$.

\section*{Appendix B}

The following R function estimates the model parameters $p$, $\beta$
and $\alpha$ through an EM algorithm.

\begin{verbatim}
fit.WG<-function(x,par,tol=1e-4,maxi=100){

# x       Numerical vector of data.

# par Vector of initial values for the parameters p, beta and alpha
# to be optimized over, on this exactly order.

# tol Convergence tolerance.

#maxi Upper end point of the interval to be searched.

p<-par[1]
beta<-par[2]
alpha<-par[3]
n<-length(x)
z.temp<-function(){
(1+p*exp(-(beta*x)^alpha))/(1-p*exp(-(beta*x)^alpha))
}
alpha.sc<-function(alpha){
n/alpha+sum(log(x))-n*sum(z*x^alpha*log(x))/sum(z*x^alpha)
}
test<-1
while(test>tol){
z<-z.temp()
alpha.new<-(alpha.sc,interval=c(0,maxi))$root
beta.new<-(n/sum(x^alpha.new*z))^(1/alpha.new)
p.new<-1-n/sum(z)
test<-max(abs(c(((alpha.new-alpha)),
                ((beta.new-beta)),
                ((p.new-p)))))
alpha<-alpha.new
beta<-beta.new
p<-p.new
}
c(p,beta,alpha)
}

\end{verbatim}



\end{document}